\documentclass[modern]{aastex62}
\usepackage{amssymb, amsmath}
\usepackage{color}
\usepackage{morefloats}

\shorttitle{QSOs Behind M31 and M33}
\shortauthors{Massey et al.}

\begin{document}

\title{The Discovery of QSOs Behind M31 and M33}

\author{Philip Massey}
\affiliation{Lowell Observatory, 1400 W Mars Hill Road, Flagstaff, AZ 86001, USA}
\affiliation{Department of Physics and Astronomy, Northern Arizona University, Flagstaff, AZ 86011-6010, USA}
\email{phil.massey@lowell.edu}

\author{Kathryn F. Neugent}
\affiliation{Department of Astronomy, University of Washington, Seattle, WA, 98195, USA}
\affiliation{Lowell Observatory, 1400 W Mars Hill Road, Flagstaff, AZ 86001, USA}
\email{kneugent@uw.edu}

\author{Emily M. Levesque}
\affiliation{Department of Astronomy, University of Washington, Seattle, WA, 98195, USA}
\email{emsque@uw.edu}

\begin{abstract}

We report the discovery of 11 newly found quasars behind the stellar disks of the spiral galaxies M31 and M33 in the fields covered by the Local Group Galaxy Survey. Their redshifts range from 0.37 to 2.15.  Most are X-ray, UV, and IR sources. We also report the discovery of 5 normal background galaxies.   Most of these objects were observed owing to their anomalous colors, as part of a program (reported elsewhere) to confirm spectroscopically candidate red supergiant plus B star binaries; others were discovered as part of our identification of early-type massive stars based upon their optical colors.  There are 15 previously known quasars in the same fields, for a grand total of 26,  15 behind M31 and 11 behind M33.   Of these, only eight were discovered as part of surveys for quasars; the rest were found accidentally.  The quasars are well distributed in the M31 and M33 fields, except for the inner regions, and have the potential for being good probes of the interstellar medium in these stellar disks, as well as serving as zero-point calibrators for {\it Gaia} parallaxes.
\vskip -10pt
\end{abstract}


\section{Introduction}

Foreground contamination has proven to be a principal bane of stellar population studies of nearby galaxies.  For instance,  identifying the yellow supergiant population of the Magellanic Clouds \citep{NeugentSMC, NeugentLMC} and in the Local Group spirals M31 and M33 \citep{DroutM31, DroutM33} required spectroscopy of thousands of stars to pick out these rare objects amidst the sea of Milky Way yellow dwarfs as these stars are otherwise inseparable in the pre-{\it Gaia} era.    However, our studies of the massive star populations of such systems has also revealed the occasional {\it background} object, such as the five quasars found as part of a search for Wolf-Rayet (WR) stars in M33 \citep{NeugentM33} and M31 \citep{NeugentM31}; their redshifts placed a strong emission line in our C\,{\sc iii} $\lambda 4650$ and/or He\,{\sc ii} $\lambda 4686$ detection filters.   

The identification of quasars and their ilk behind the stellar disks of nearby spirals is useful for a variety of purposes.  First, these sources
allow us to probe the interstellar medium of these stellar disks.  Hot luminous stars allow us to probe the shape of the reddening law 
and the composition of the interstellar medium (ISM) along the lines-of-sight to regions of recent star formation in the where such stars abound, but background QSOs allow us to measure the total extinction through the stellar disk as a function of position.  (For a recent applications, see \citealt{Petko}.)  Furthermore, QSOs located (in projection) in the outer parts of the stellar disks allow us to probe the ISM of the halo gas.  Others have used quasars to study the properties of the extended H\,{\sc i} disk of M31 (see, e.g., \citealt{2013MNRAS.432..866R}), and the circumgalactic medium surrounding M31 (see, e.g., \citealt{Lehner}, and \citealt{Howk}.)

Secondly,  background point sources provide a useful calibrant for {\it Gaia} parallaxes.  Our studies of the massive star populations of Local Group galaxies is being revolutionized by the availability of Data Release 2 (DR2) {\it Gaia} parallaxes and proper motions \citep{DR2}, making it relatively straightforward to weed out foreground Milky Way disk stars from our samples.  (See, e.g., \citealt{ErinLBV}.)  However, the separation of foreground halo objects (potentially at distances of 5-10~kpc) from more distant objects, such as M31/M33 stars at $\sim$800~kpc, is a lot trickier, as the {\it Gaia} parallax zero-points in DR2 suffer from systematic errors on the order of 0.03~mas with the value dependent upon position in the sky \citep{Lindegren}.  One approach is to use a large number of {\it Gaia} measurements of the stars in the host
galaxy to define a median.  However, even objects at 1~Mpc (the fringes of the Local Group) will have a parallax of 1 micro-arcsec, within the ultimate expected precision of {\it Gaia}.  It is thus useful to have point sources that are truly at ``infinity" for all practical purposes to serve as a uniform set of references.

During a recent search for red supergiant (RSG) binaries in M31 and M33, we obtained spectra of targets selected from the Local Group Galaxy Survey (LGGS, \citealt{LGGSII,BigTable}) based upon their having ``composite" colors, i.e., {\it U-B} values typical of B stars but {\it R-I} colors typical of RSGs, following
the prescription of \citet{NeugentRSGBinI}.   The major results of that survey are described elsewhere \citep{NeugentRSGBinII}, but in the course of that study we discovered 5 additional background QSOs as well as several background galaxies.   Our spectroscopic surveys for hot stars in M31 and M33 over the years had discovered other QSOs that we included in the revised LGGS catalog published as part of \citet{BigTable}, but we did not call attention to these nor provide redshifts.  Given our new discoveries, it seemed the right time to rectify this oversight.

\section{Observations and Reductions}
All observations reported here were taken with the Hectospec mutiobject fiber-fed spectrometer \citep{Hecto} on the 6.5-m MMT telescope at the Fred Whipple Observatory on Mount Hopkins.  We list the observational details in Table~\ref{tab:journal}.   A little over half of the new data were
taken during Fall 2018 using the 270 line mm$^{-1}$ grating, which yielded a spectral resolution of 5~\AA, and covered the entire optical
region from 3650~\AA\ to 9200~\AA.  Earlier data were mostly obtained with the 600 line mm$^{-1}$ with a spectral resolution of 2~\AA\, and coverage from
3700\AA\ to 6000\AA.   We list the exposure times and grating setups in Table~\ref{tab:journal} in an object-by-object order for the data discussed in this paper, but these were not separate exposures: data taken on the same night were taken at the same time as part of the same fiber configuration.  The data were obtained as part of the Hectospec queue, but for nearly all the observations the first two authors were present and took the actual exposures. The 2018 data were reduced using the Version 2.0 of the {\sc hsred idl} package (see \url{https://www.mmto.org/node/536}); the older data had been reduced by the {\sc specroad iraf} pipeline.  

Data taken with the 270 line mm$^{-1}$ grating were flux calibrated using calibration files kindly provided by Nelson Caldwell no spectrophotometric standards were observed in the queue with the 600 line mm$^{-1}$ grating, and we have left those spectra in units of raw counts rather than attempt normalization.  We remind the reader that these objects have broadband optical photometry {\it UBVRI} as part of the LGGS. 
 
\section{QSOs Behind M31 and M33}

\subsection{New Discoveries and Redshifts}

Although the term ``quasar" originally referred to quasi-stellar {\it radio} sources, it is now recognized that only about 5-10\% of quasars are radio strong, and the term is now used interchangably with ``QSO" (quasi-stellar object); for more on the taxonomy of active galactic nuclei (AGN), the reader is referred to \citet{Peterson}.  Our recognition of a QSO is based upon the presence of strong, broad emission lines UV (typically L$\alpha$,  Si\,{\sc iv}+O\,{\sc iv} $\lambda 1400$, C\,{\sc iv} $\lambda 1549$,  He\,{\sc ii} $\lambda 1640$, C\,{\sc iii]} $\lambda 1909$, Mg\,{\sc ii} $\lambda 2799$) which have been redshifted into the optical. We believe that these are all what would be called the classical type-I broad-lined QSOs.  As shown below, all of the objects we are calling QSOs are point sources on the LGGS images, and therefore cannot be called Seyfert I galaxies despite the similarity type-I QSOs have in spectral appearance (see, e.g., \citealt{Peterson}). 

The eleven newly discovered QSOs are listed in Table~\ref{tab:QSOs} along with their redshifts; their spectra are shown in Figure~\ref{fig:qsos}.
In determining the line identifications we compared the observed wavelengths against the SDSS quasar emission line list \url{http://classic.sdss.org/dr6/algorithms/linestable.html}.  Agreement within a redshift $z$ of 0.01 between multiple lines was taken as an indication that we had the right match and the correct $z$;
the lines that showed such agreement were invariably what we expect to be the strongest quasar lines.

We have checked the NASA/IPAC Extragalactic Database (NED), VizieR, and Simbad for all of these sources, and included any further pertinent information under comments in Table~\ref{tab:QSOs}.  Note that all of these objects are bright across multi-wavelength space, i.e., from the X-ray through the IR.  This is typical of quasars in general.  In Table~\ref{tab:QSOs} we have listed only a single mission associated with the detection in the interest of space; typically there are detections in the X-ray and IR from multiple sources, as can be found in the on-line databases mentioned above.

\begin{figure}
\epsscale{0.45}
\plotone{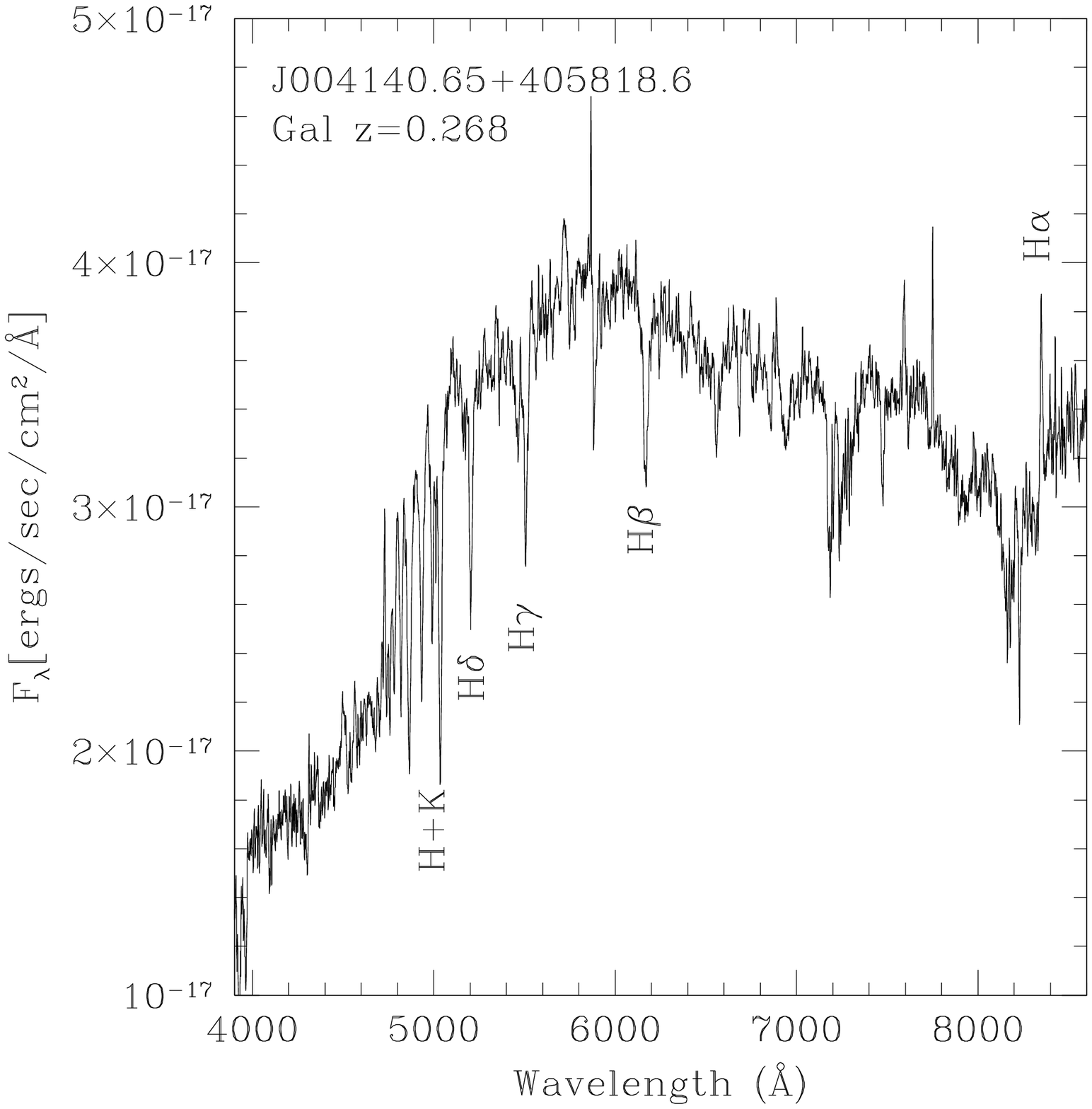}
\plotone{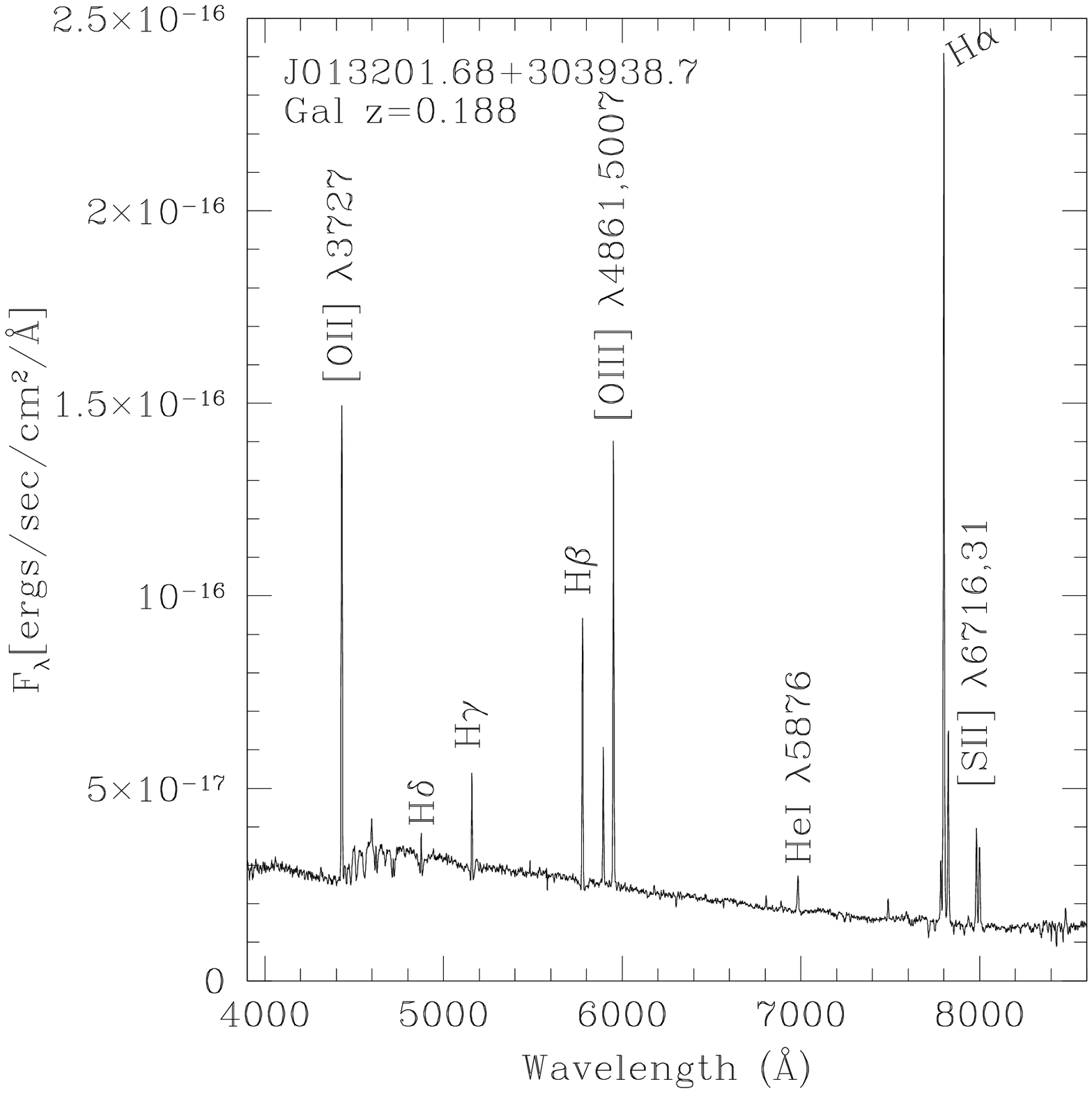}
\plotone{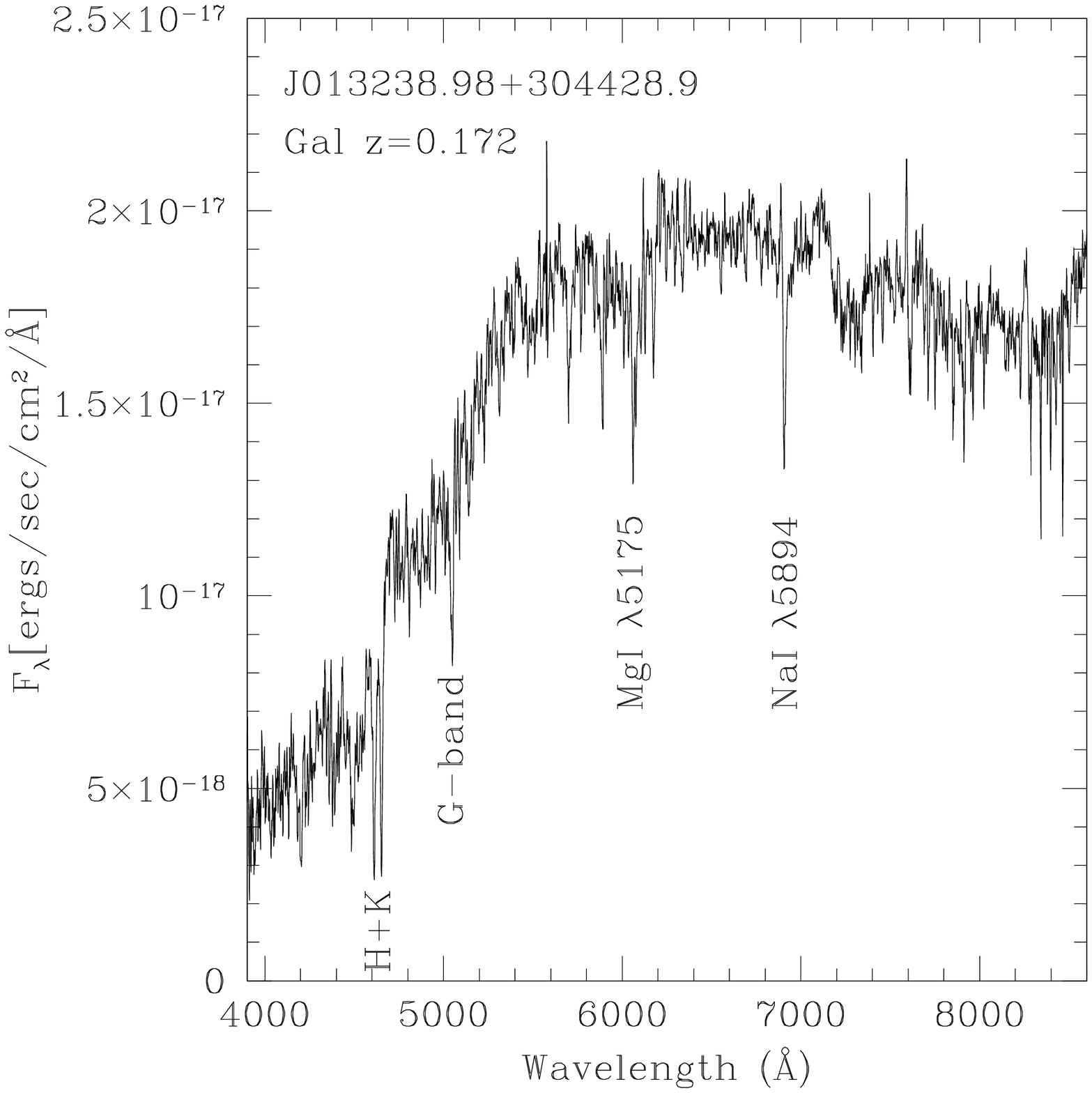}
\plotone{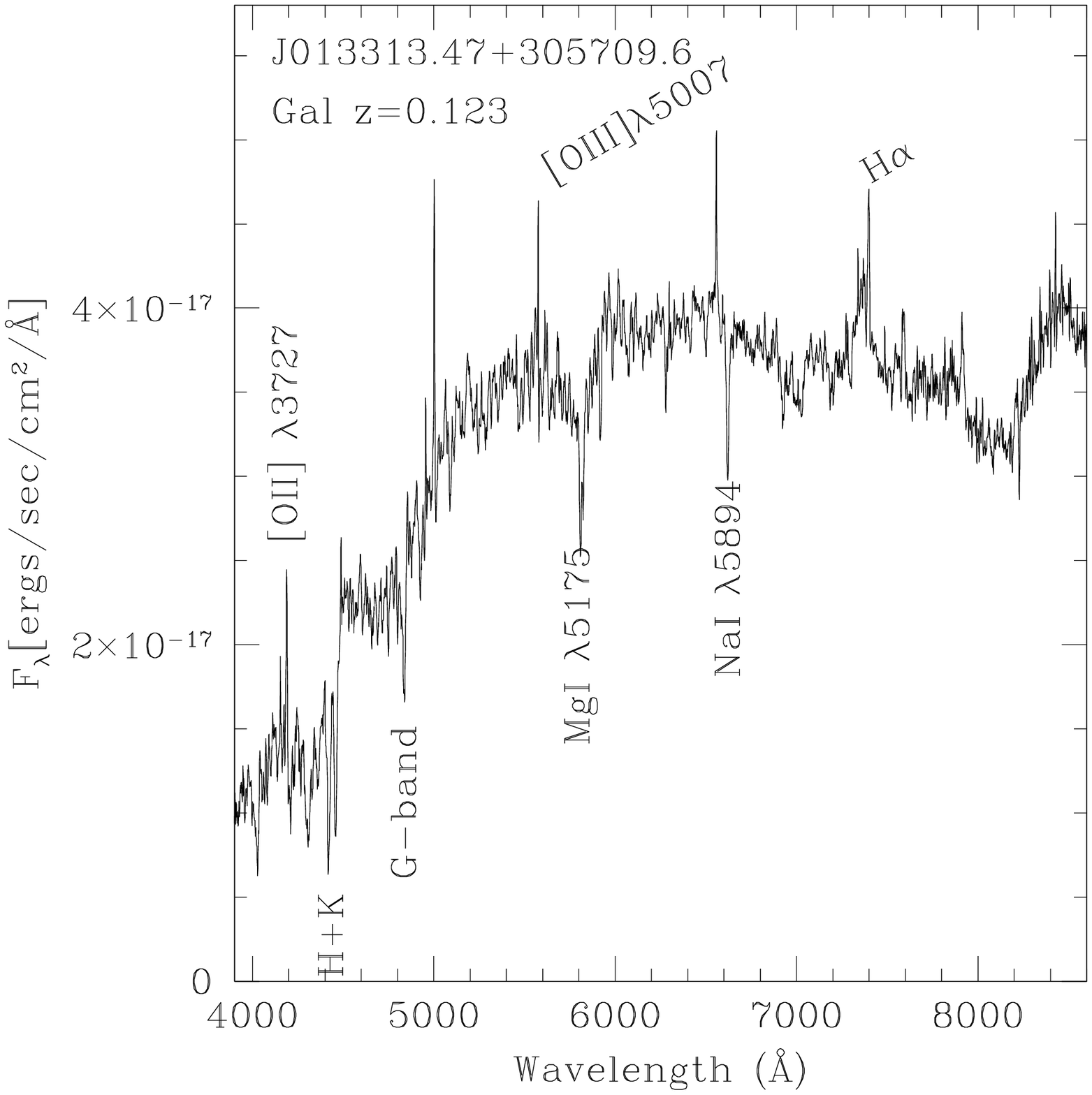}
\plotone{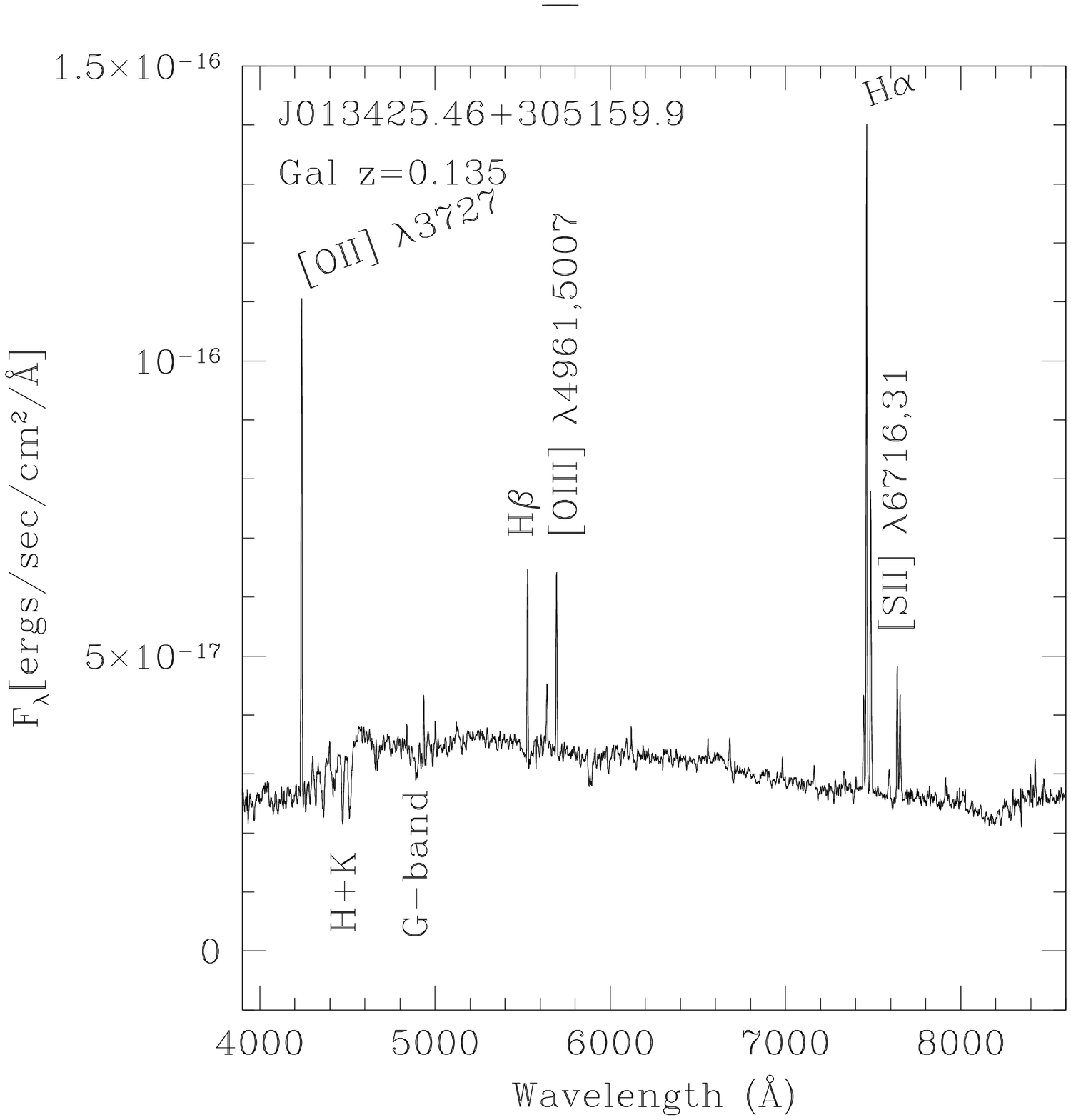}
\caption{\label{fig:gals} Spectra of Newly Found Galaxies from Table~\ref{tab:QSOs}.  The prominent spectral lines are identified.	
}
\end{figure}

Along with these eleven quasars, we also discovered five previously unknown background galaxies, one behind M31 and four behind M33.  These all have absorption spectra, sometimes with nebular emission; the latter is indicative of recent star formation.  We include these in lower part of Table~\ref{tab:QSOs}, and show their spectra in Figure~\ref{fig:gals}.  Four had been considered globular cluster candidates; one had been ironically misidentified as a probable foreground red dwarf in the ``control field" of a search for RSGs in nearby galaxies by the first author of the present paper.  In some cases the absorption spectrum resembled were characteristic of A or F-type stars, with strong Balmer absorption and the presence of the G-band; in others the spectrum resembled what we would expect from a much older population, such as what would be found in an elliptical galaxy.

\subsection{Previous Known QSOs}

Prior to the present study, only 15 other quasars were known within the LGGS M31 and M33 fields, as listed
in Table~\ref{tab:Previous}.  It is worth briefly reviewing how these were discovered.

As mentioned above, five of the previously known QSOs were discovered by \citet{NeugentM33} and \citet{NeugentM31} as part of surveys for WR stars in M33 and M31, respectively.  The redshifts of these QSOs just happened to place one of their strong lines within the bandpass of our C\,{\sc iii} $\lambda 4650$ and/or He\,{\sc ii} $\lambda 4686$ filters.  One (J013322.09+301651.4) had actually been mis-classified as a WR star in an earlier study (\citealt{MC83}; see discussion in \citealt{NeugentM33}).

Two of the other 11 QSOs were found as follow ups to what appeared to be peculiar objects. J004457.94+412343.9 was originally thought to be a peculiar nova (object 21 in \citealt{Sharov}); its nature as a QSO was identified by \citet{2010AA...512A...1M}, where is referred to as ``J004457+4123."  This is the object used by \citet{Petko} as a probe of M31's ISM.  We have matched it to the LGGS by precessing \citet{Sharov} very precise position.  As for J004527.30+413254.3,  \citet{Fred} were searching for red supergiant X-ray binaries, and this source came to their attention as its colors suggested it was a RSG but it was also in the Chandra Source Catalog as an X-ray source; the only object to share both characteristics.  Their spectrum, however, showed redshifted broad emission features.  (Previously the object had been
called a  globular cluster by \citealt{2007AJ....134..706K}, and later it was (mis)identified as an eclipsing binary by \citealt{Vilardell}.) 

Three others were identified as quasars from among the four million spectra obtained as part of the spectroscopic followup for the Sloan Digital Sky Survey \citep{2017AA...597A..79P}.  It may be worth noting that
none of the other 23 QSOs in our fields were identified in this way.

All of the other five were discovered as part of directed searches for QSOs behind M31 and M33. 
The quasar nature of J004528.26+412944.2  was reported briefly recently in \citep{2018ATel12250....1N} as part of a spectroscopic followup of QSO candidates identified from mid-IR photometry.  P. Nedialkov (2019, private communication) informs us they will publish their identifications and redshifts of several others shortly, and they have plans to continue their search.   J004655.52+422050.2 is one of the ten QSOs discovered behind M31 in a search specifically to study the ISM by \citet{2013MNRAS.432..866R}; however,  it is the only one coincident with an LGGS source, with the others located further from the stellar disk. Finally,  three were discovered as  part of the LAMOST survey for QSOs \citep{LAMOST} behind M31 and M33.  Again, note that the other 23 QSOs in our fields were not detected by that effort.

\section{Discussion and Conclusions}

\begin{figure}
\epsscale{1.5}
\plotone{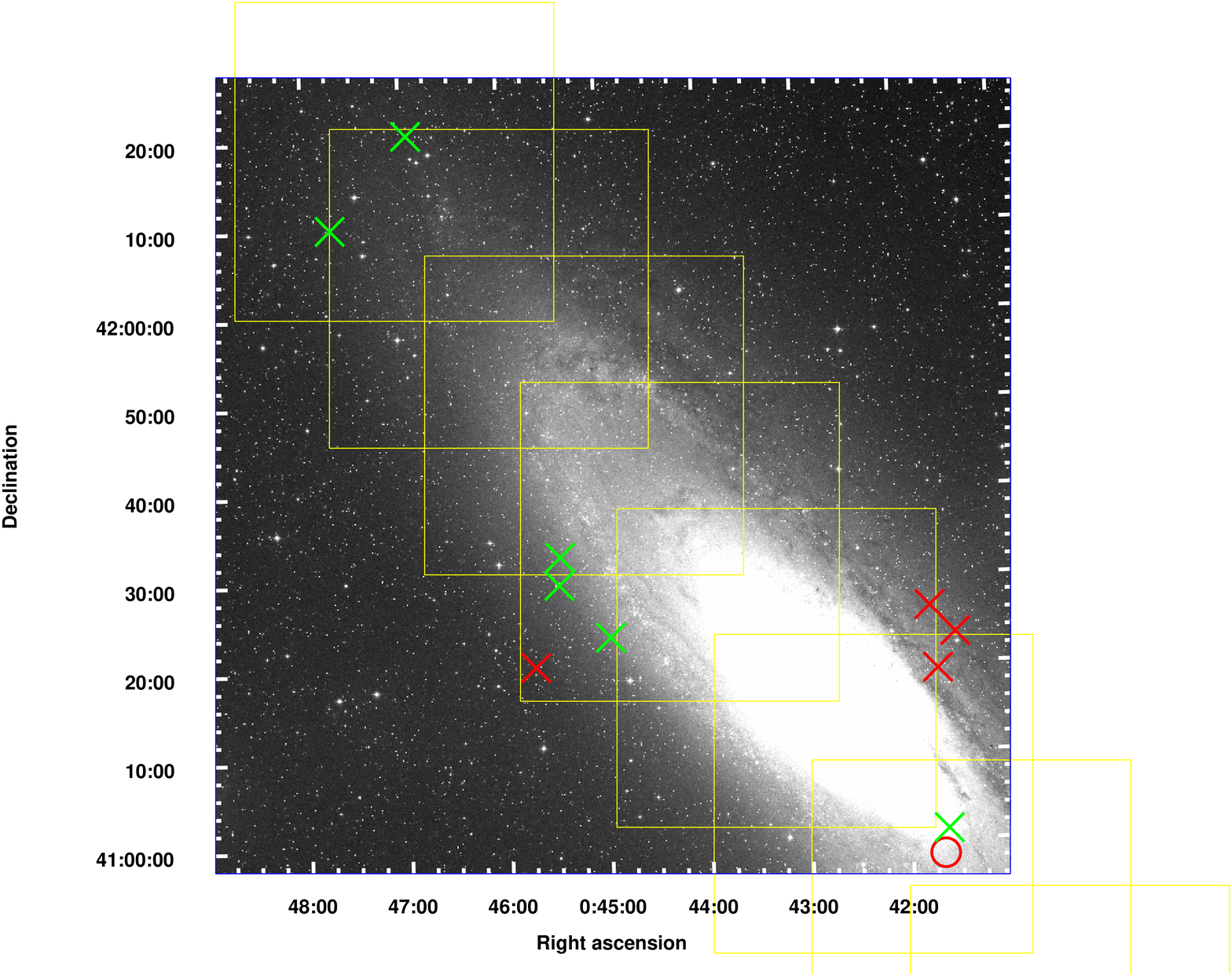}
\caption{\label{fig:m31NE}. Location of M31 Objects (NE field).   The red x's show the location of the newly found QSOs, while the green x's denote the previously known QSOs.  The red circles show the locations of the newly found galaxies.  The image is from the red Digital Sky Survey, and is 1.5$^\circ$ on a side. The LGGS fields are outlined in yellow.
}
\end{figure}

\begin{figure}
\epsscale{1.5}
\plotone{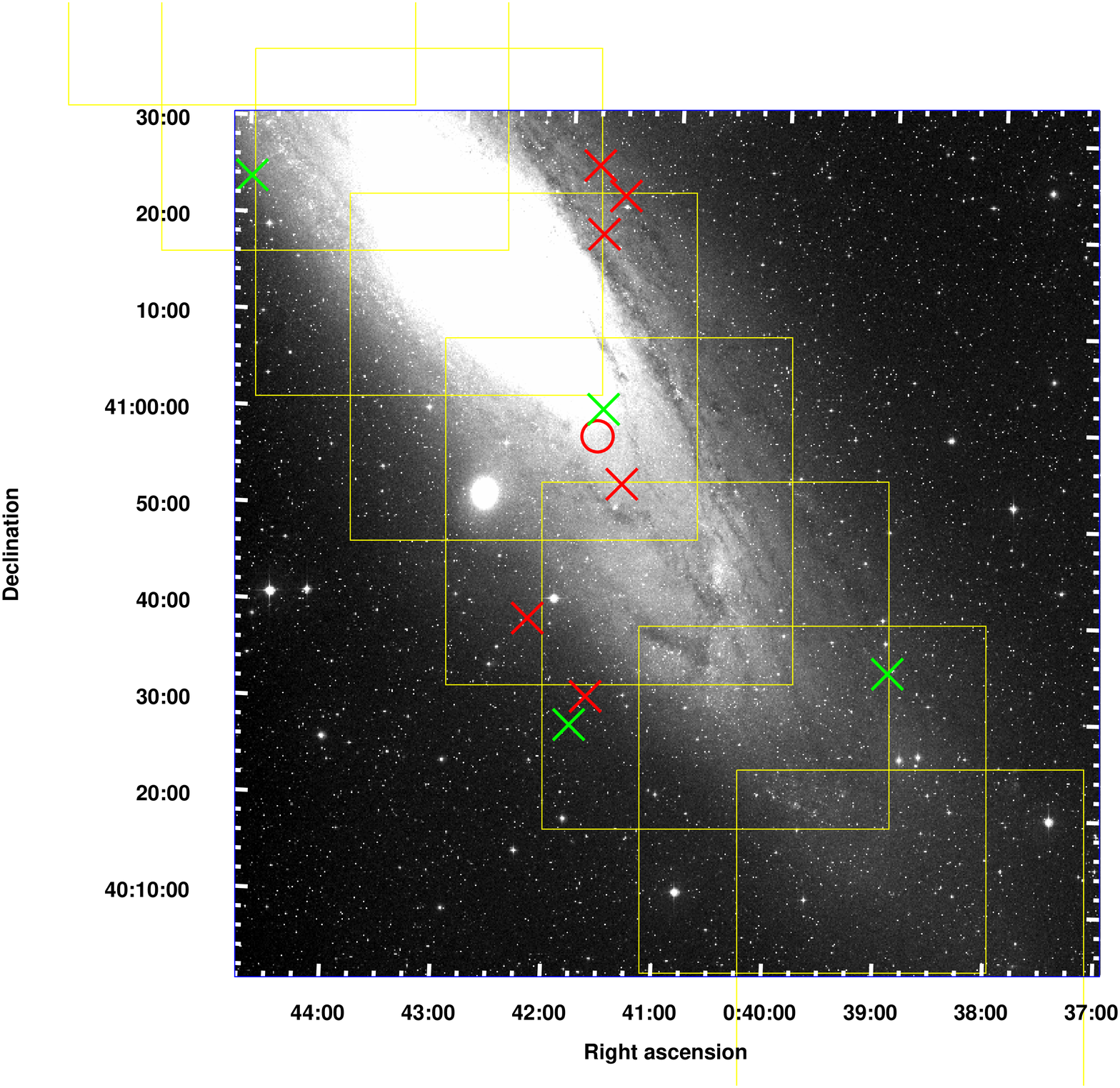}
\caption{\label{fig:m31SW}. Location of M31 Objects (SW field).   The red x's show the location of the newly found QSOs, while the green x's denote the previously known QSOs.  The red circles show the locations of the newly found galaxies.  The image is from the red Digital Sky Survey, and is 1.5$^\circ$ on a side. The LGGS fields are outlined in yellow.
}
\end{figure}

\begin{figure}
\epsscale{1.5}
\plotone{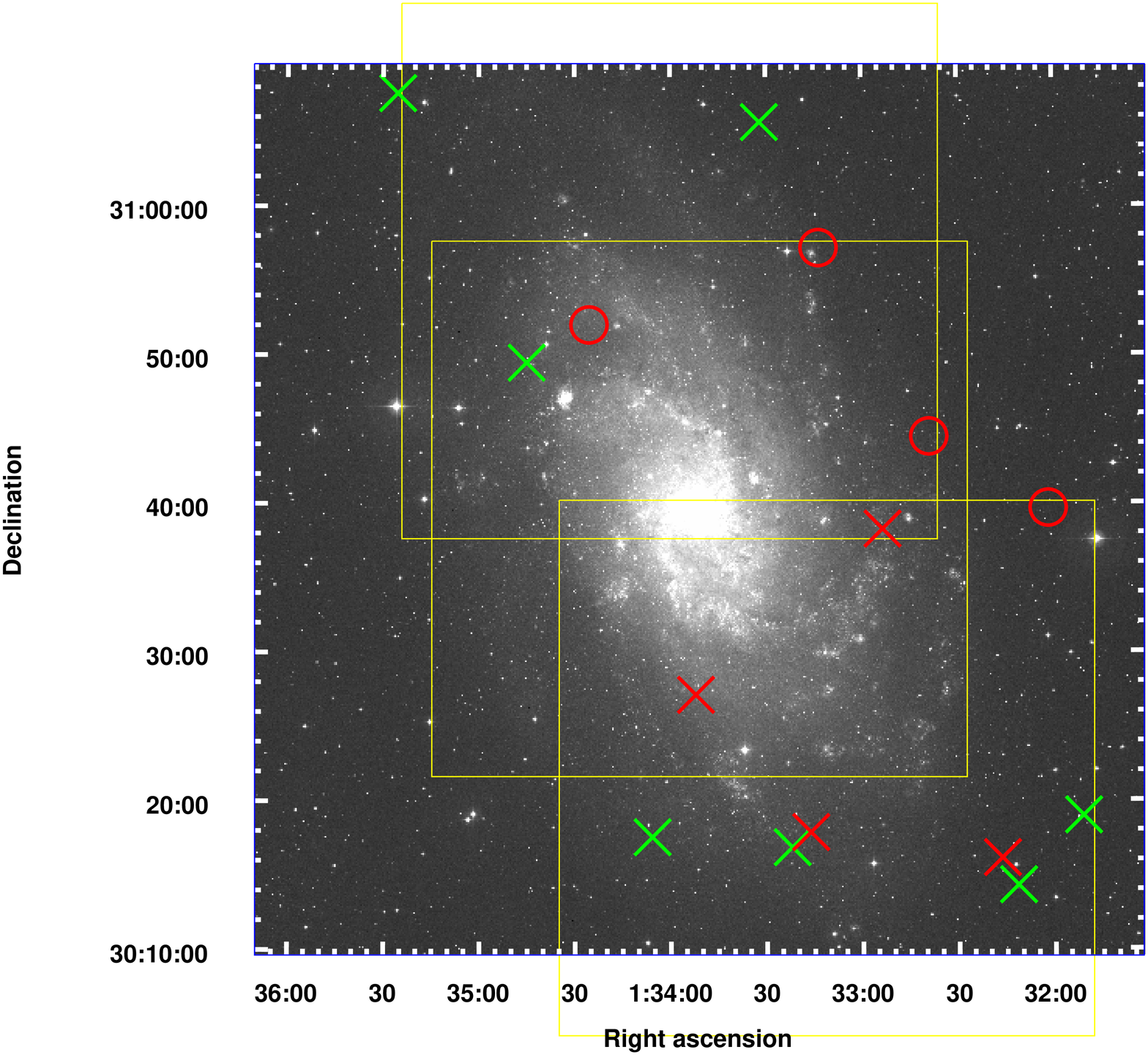}
\caption{\label{fig:m33}. Location of M33 Objects.  The red x's show the location of the newly found QSOs, while the green x's denote the previously known QSOs.  The red circles show the locations of the newly found galaxies.  The image is from the red Digital Sky Survey, and is 1$^\circ$ on a side. The LGGS fields are outlined in yellow.
}
\end{figure}

In Figures~\ref{fig:m31NE}-\ref{fig:m31SW} and \ref{fig:m33} we show the locations of the known LGGS quasars and newly found galaxies in M31 and M33, respectively.   Note that in both galaxies there is good coverage throughout the stellar disk, with the exception of the bulge of M31 and the central-most part of M33. Our discoveries have better populated the central regions of both galaxies. As shown by the galactocentric $\rho$ values in Tables \ref{tab:QSOs} and \ref{tab:Previous}, these quasars can be used to sample the stellar disks of these
galaxies from Holmberg radii of  0.2 to 1.2 for M31, and from 0.5-1.3 for M33.

We  have examined each of of the objects on the $V$-band LGGS frames, comparing their image shapes to neighboring stars; without exception, all of the quasars listed here (both newly found and previous known) are point-like on these images. (The images sizes were typically $<$1\arcsec.) The objects with galaxy spectra are all slightly extended, typically by a few tenths of a pixel. (The image scale is 0\farcs24 pixel$^{-1}$.) The exception is the galaxy J013201.68+303938.7, which is clearly extended by eye and has a full-width-half-maximum of 1\farcs0 compared to neighboring objects with image sizes of 0\farcs8.

Of our newly found 11 QSOs reported here, five were found as a result of their having anomalous colors, being both blue (in {\it U-B}) and red (in {\it R-I}) at the same time, suggestive of RSG+B binaries \citep{NeugentRSGBinII}.  Another five were found as part of our spectroscopy of hot stars in M31 and M33 \citep{BigTable}.  One was found as a follow-up to narrow-band photometry designed to detect peculiar emission line stars \citep{LGGSIII}. With the addition of these 11,  there are now 26 confirmed QSOs known behind the stellar disks of M31 and M33.  It is interesting to note that of these 26, only 8 were the result of looking for QSOs, and of those only 5 were found by efforts to find QSOs specifically behind M31 and M33.  {\it Our conclusion is that finding QSOs behind these galaxies is hard.}\footnote{At least on purpose.}

We note that in general it is important to get better at identifying background AGNs, as their variability could easily be mistaken for interesting objects within Local Group galaxies, as was the originally the case for several of the QSOs mentioned here.  In this age of transient searches,  the reverse is also true: a high luminosity gamma-ray burst (GRB) in a background object could easily be mistaken for a low-luminosity GRB in M31 itself, say.   Thus improving our knowledge of the background galaxies and quasars for these nearby galaxies is doubly useful. 

One potential way to increase their numbers is to obtain spectra of the various WISE and SDSS photometric QSO candidates. We note that three of the 11 newly confirmed QSOs in Table~\ref{tab:QSOs} were actually photometrically identified as QSO candidates in this way; furthermore, their ``photometric redshifts" were correctly inferred in two of the three cases.    It is of course not clear what the success rate would be.  M31 and M33 contain many emission-lined stars (see, e.g., \citealt{LGGSIII}).  Added value might be achieved by using variability as a further discriminant; this was used to good effect in the general region of M31 by Yuhan Yao in some unpublished work (see \url{https://speakerdeck.com/yaoyuhan/quasars-behind-m31-from-ptf-survey}). We note that one of the interesting finds we discuss in our recent RSG+B star binary paper \citep{NeugentRSGBinII}, J004032.98+404102.8 (a highly unusual high-mass symbiotic star which is a member of M31) is a WISE photometric quasar candidate with a nominal redshift of 0.7.  In addition, \citet{FRNAAS} observed two objects which they report as prime QSO candidates behind M31; neither proved to be. This is in keeping with our conclusion above that finding quasars behind these galaxies is {\it hard.}

\acknowledgements
Observations reported here were obtained at the MMT Observatory, a joint facility of the University of Arizona and the Smithsonian Institution. We are
grateful to Nelson Caldwell for his management of the Hectospec queue, and to Perry Berlind, Mike Calkins, ShiAnne Kattner, Marc Lacasse, and  Erin Martin for their invaluable assistance in obtaining the observations.  Useful comments were provided by Erin Aadland,  Petko Nedialkov, and an anonymous referee.   We are grateful to the Arizona Time Allocation Committee for their continued and generous support of our work.   This research has made use of the VizieR catalog access tool, CDS, Strasbourg, France; identification of LGGS sources with known QSOs was greatly facilitated by their {\sc XMatch} tool and the on-line compilation of a ``Million Quasars" from the literature by \citet{2017yCat.7277....0F}.   This work was supported in part
National Science Foundation (NSF) under grant
AST-1612874, and  NSF IGERT grant DGE-1258485, as well as by a fellowship from the Alfred P. Sloan
Foundation.

\facilities{MMT (Hectospec)}
\clearpage

\begin{figure}
\epsscale{0.45}
\plotone{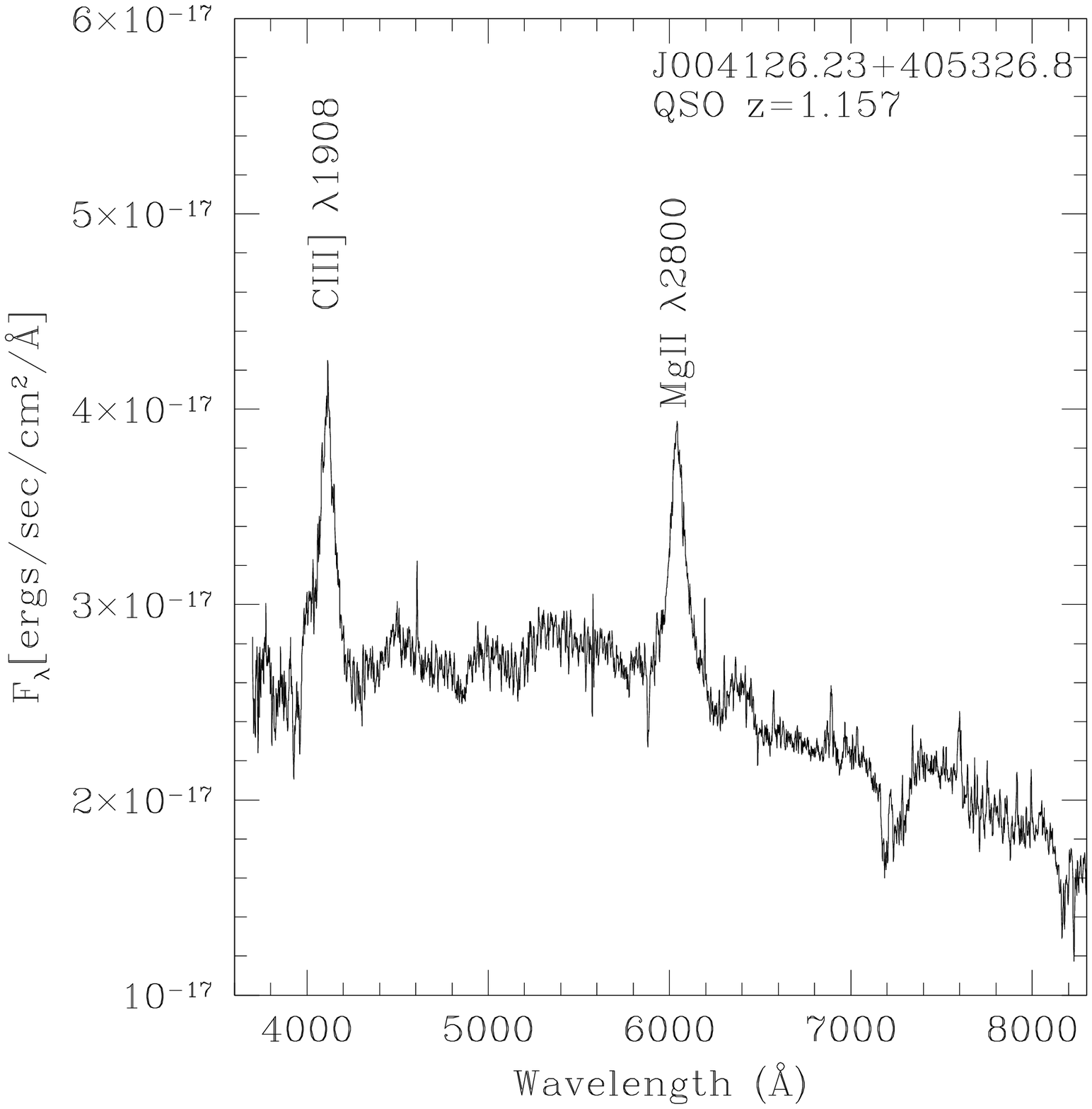}
\plotone{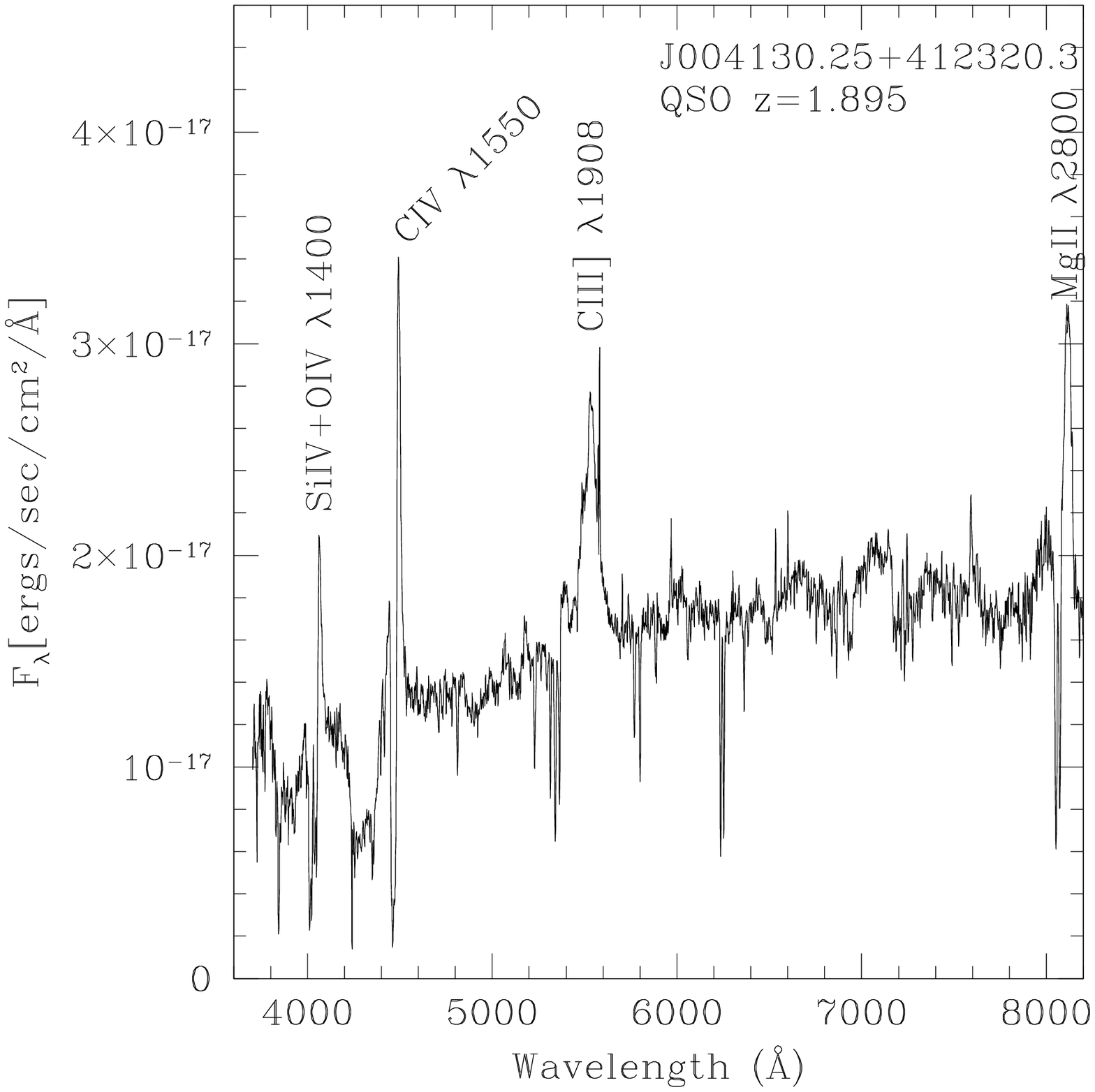}
\plotone{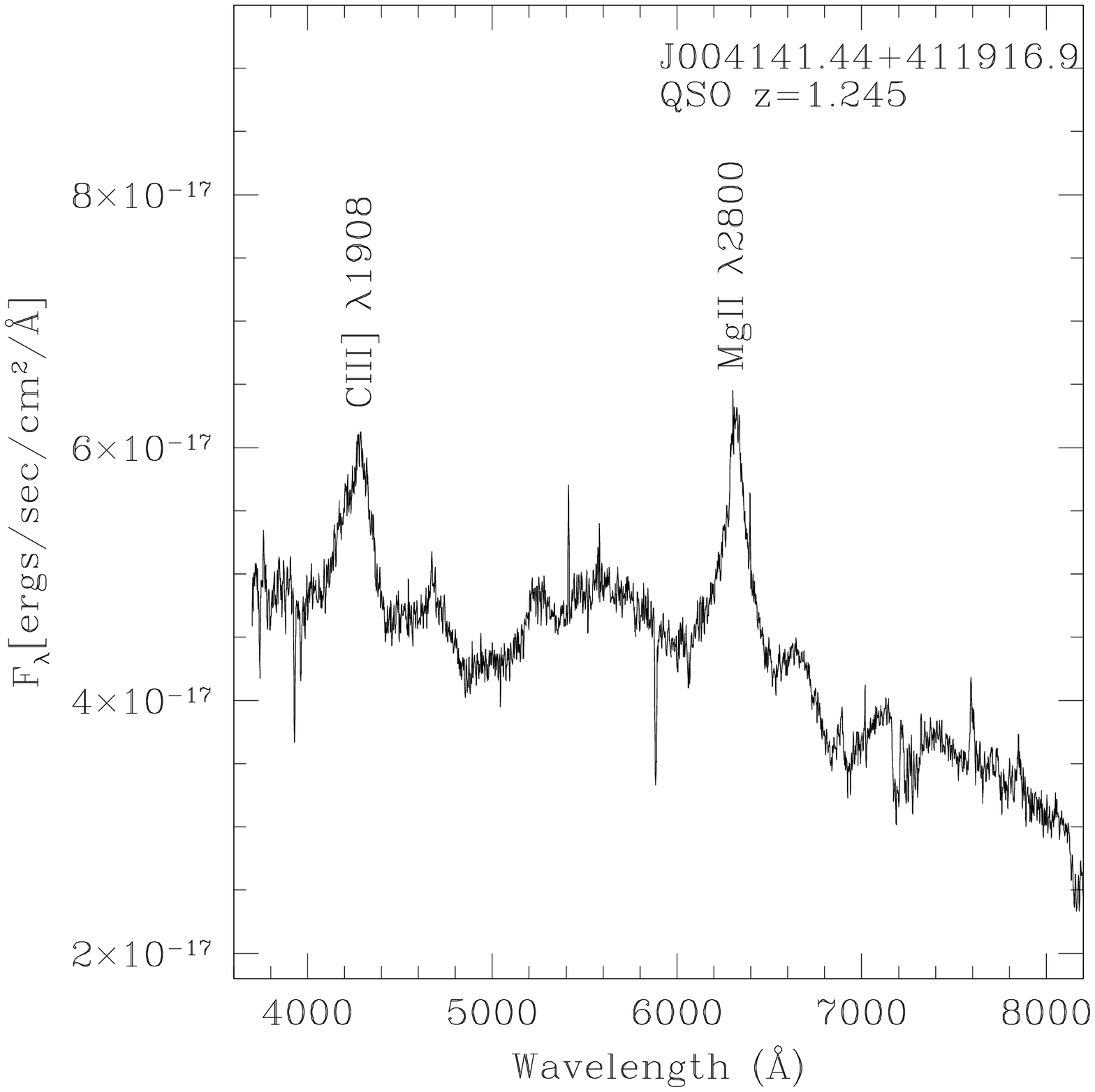}
\plotone{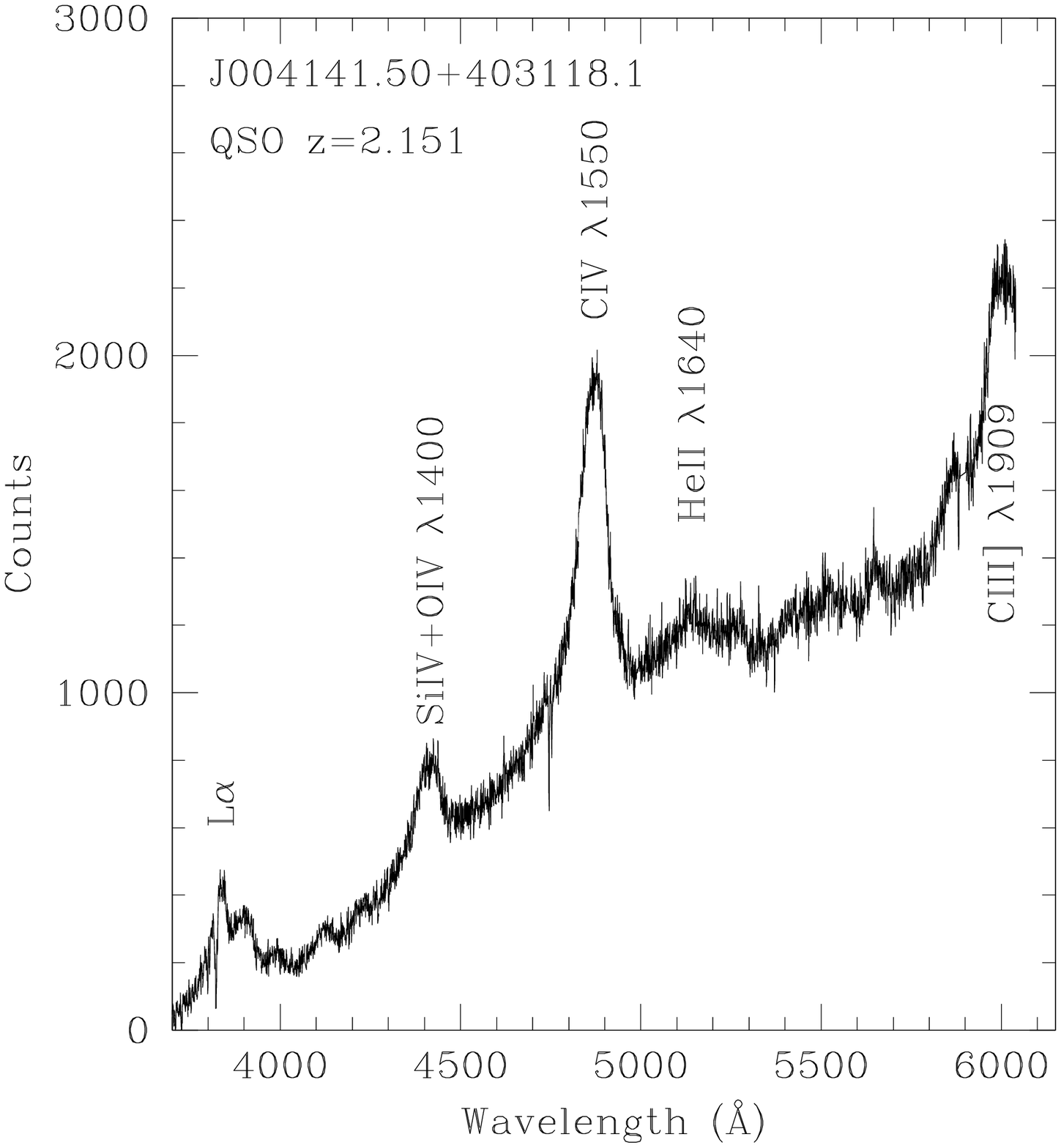}
\plotone{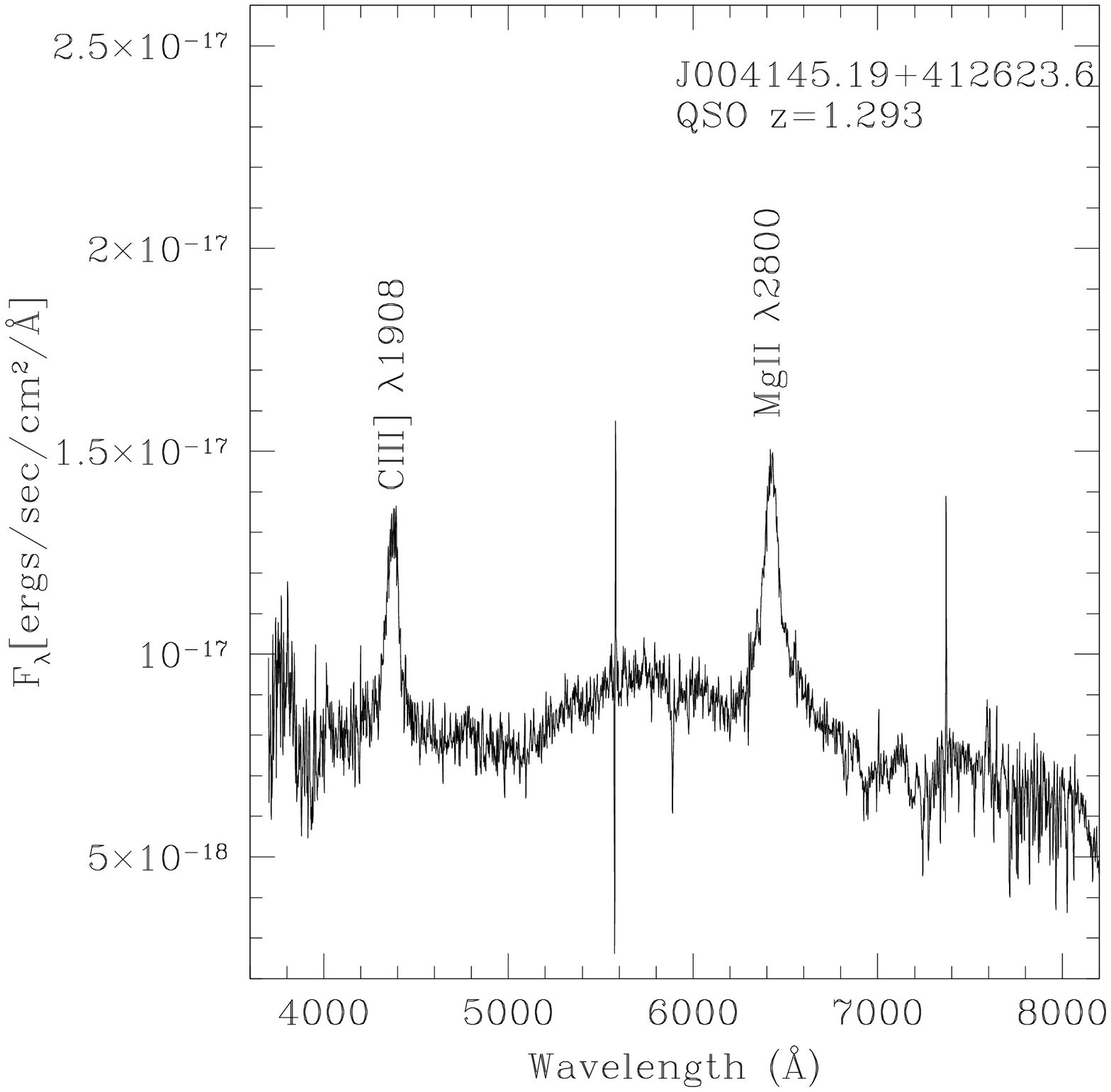}
\plotone{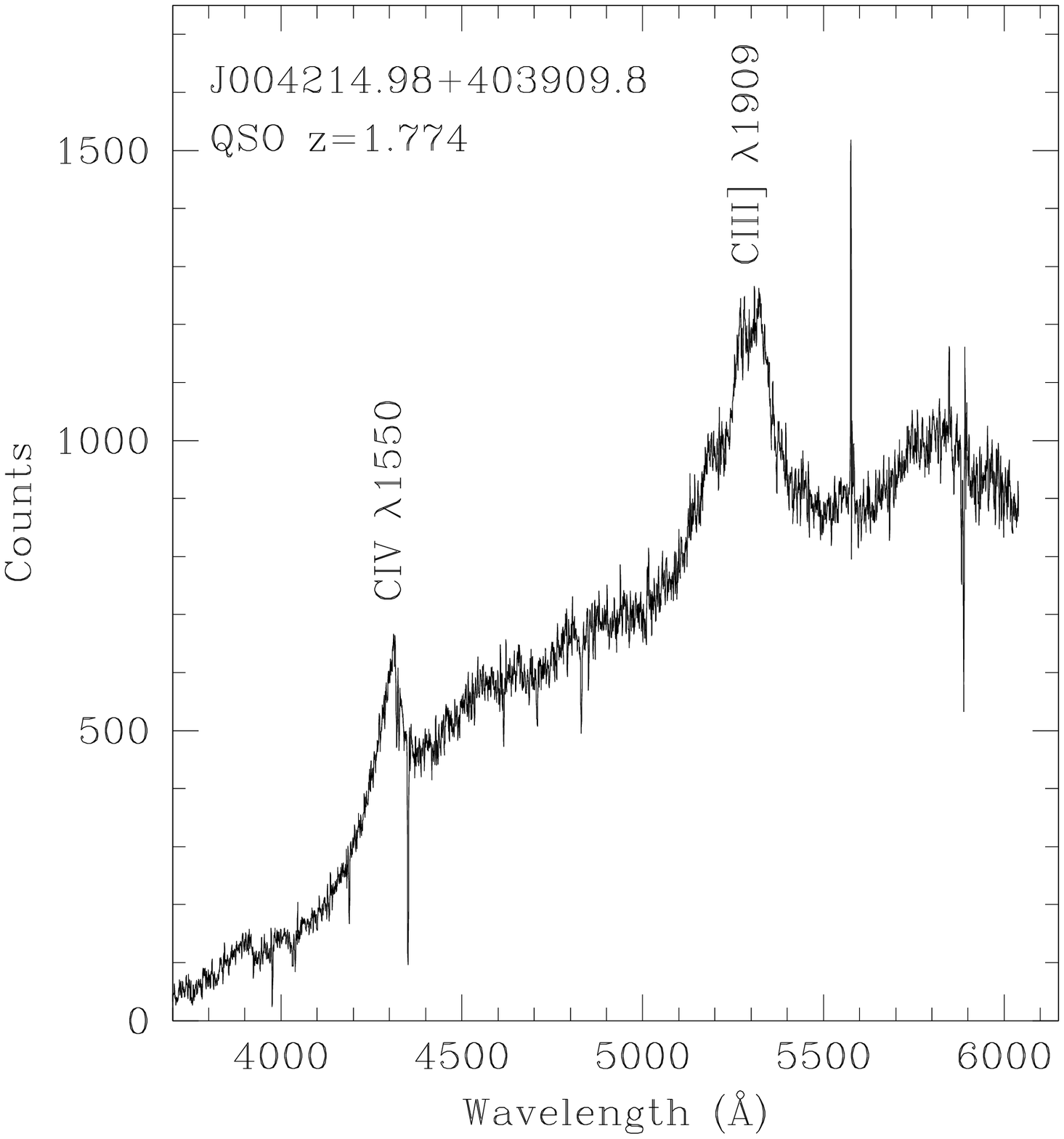}
\caption{\label{fig:qsos} Spectra of Newly Found QSOs from Table~\ref{tab:QSOs}. We have indicated the principal lines used for determining the redshifts.
}
\end{figure}

\clearpage

\begin{figure}
\figurenum{1}
\epsscale{0.45}
\plotone{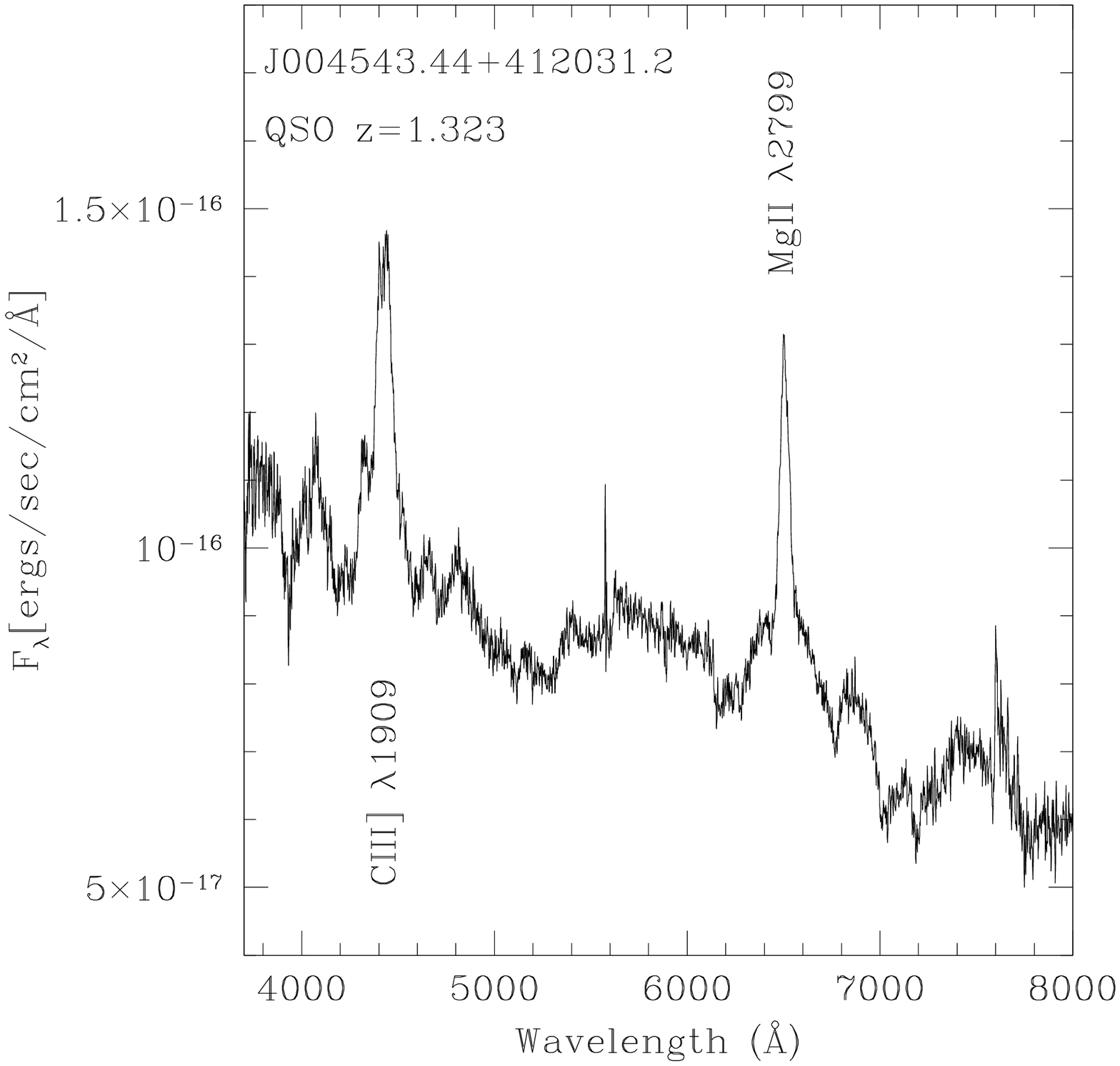}
\plotone{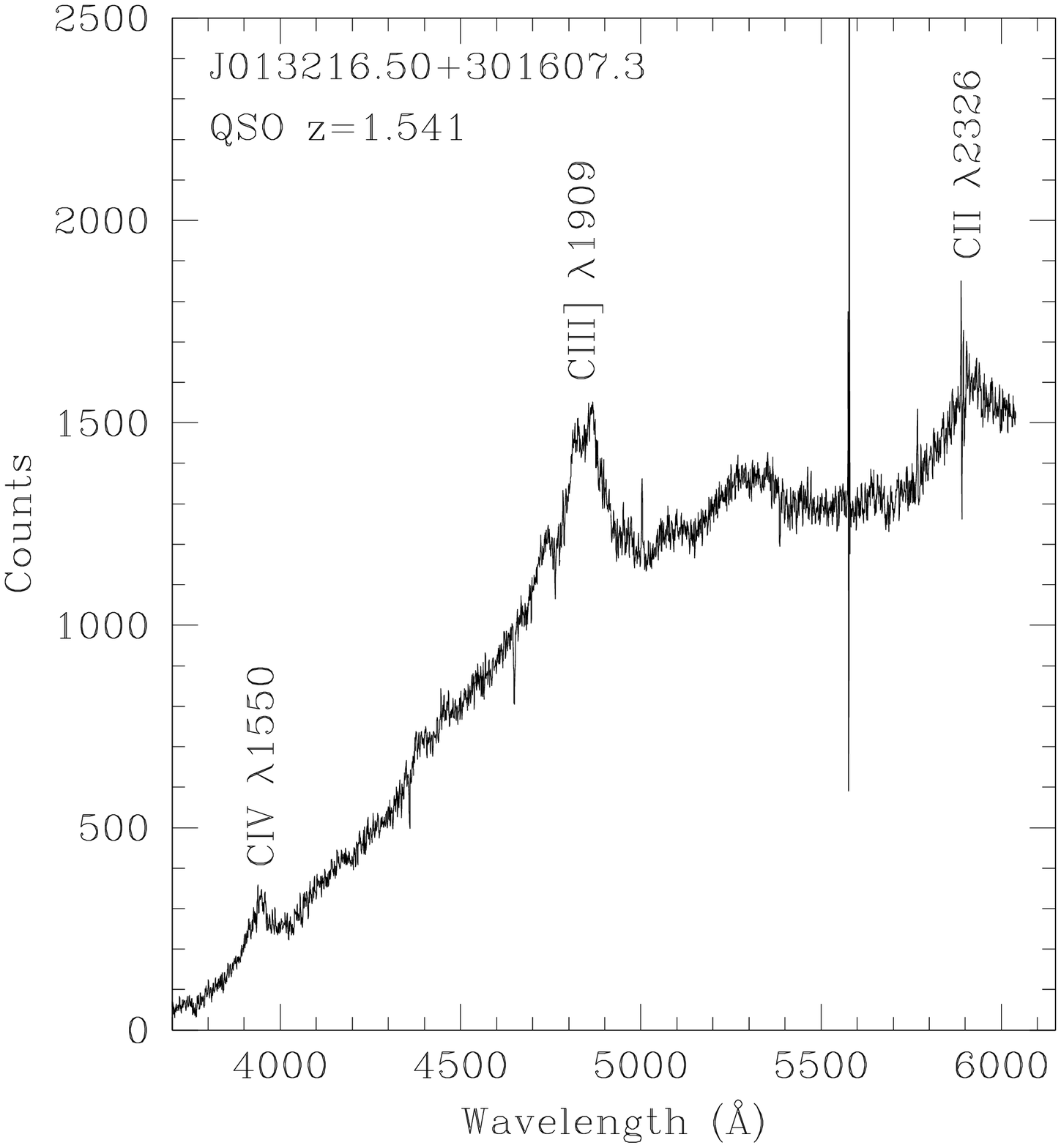}
\plotone{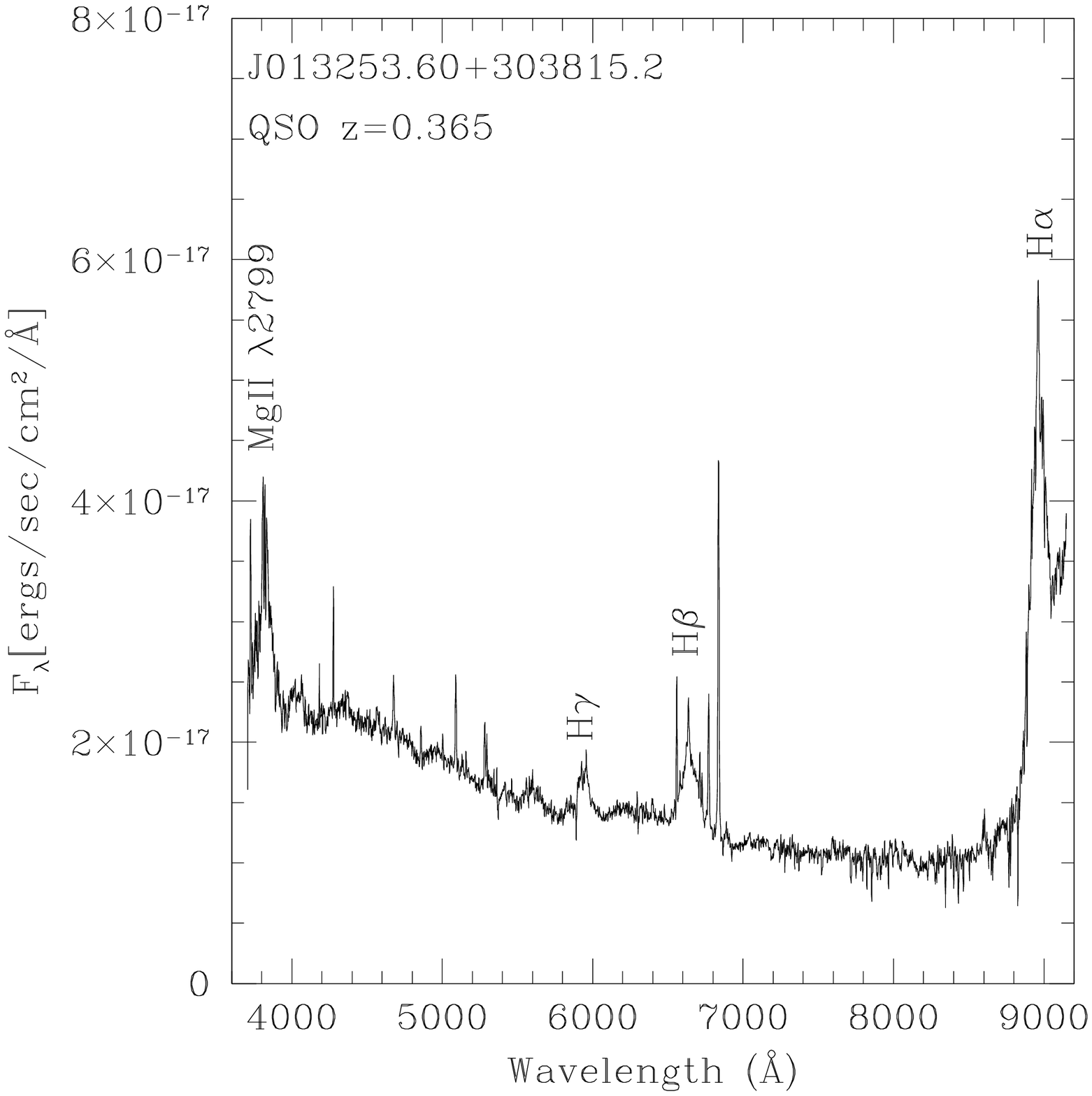}
\plotone{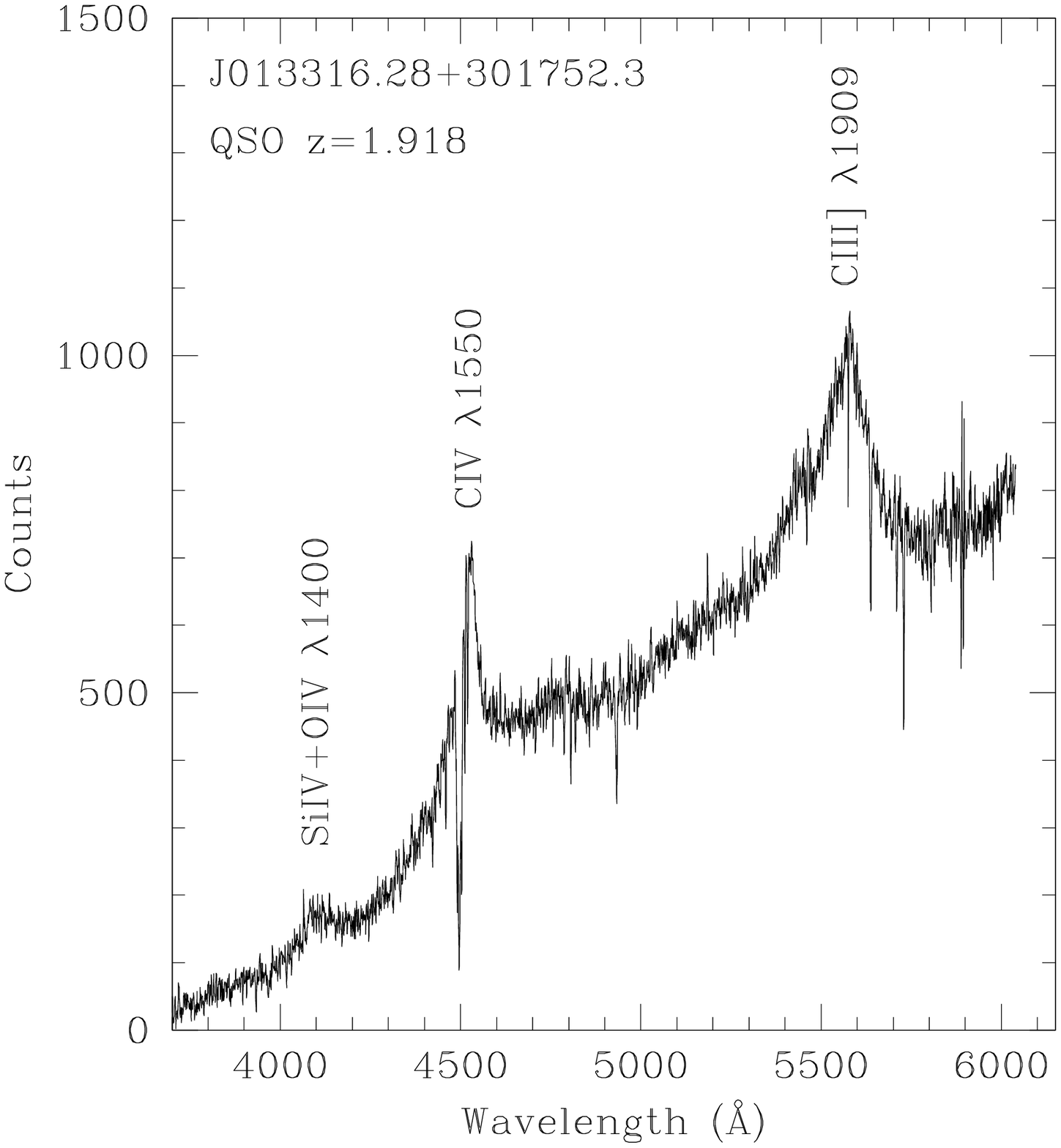}
\plotone{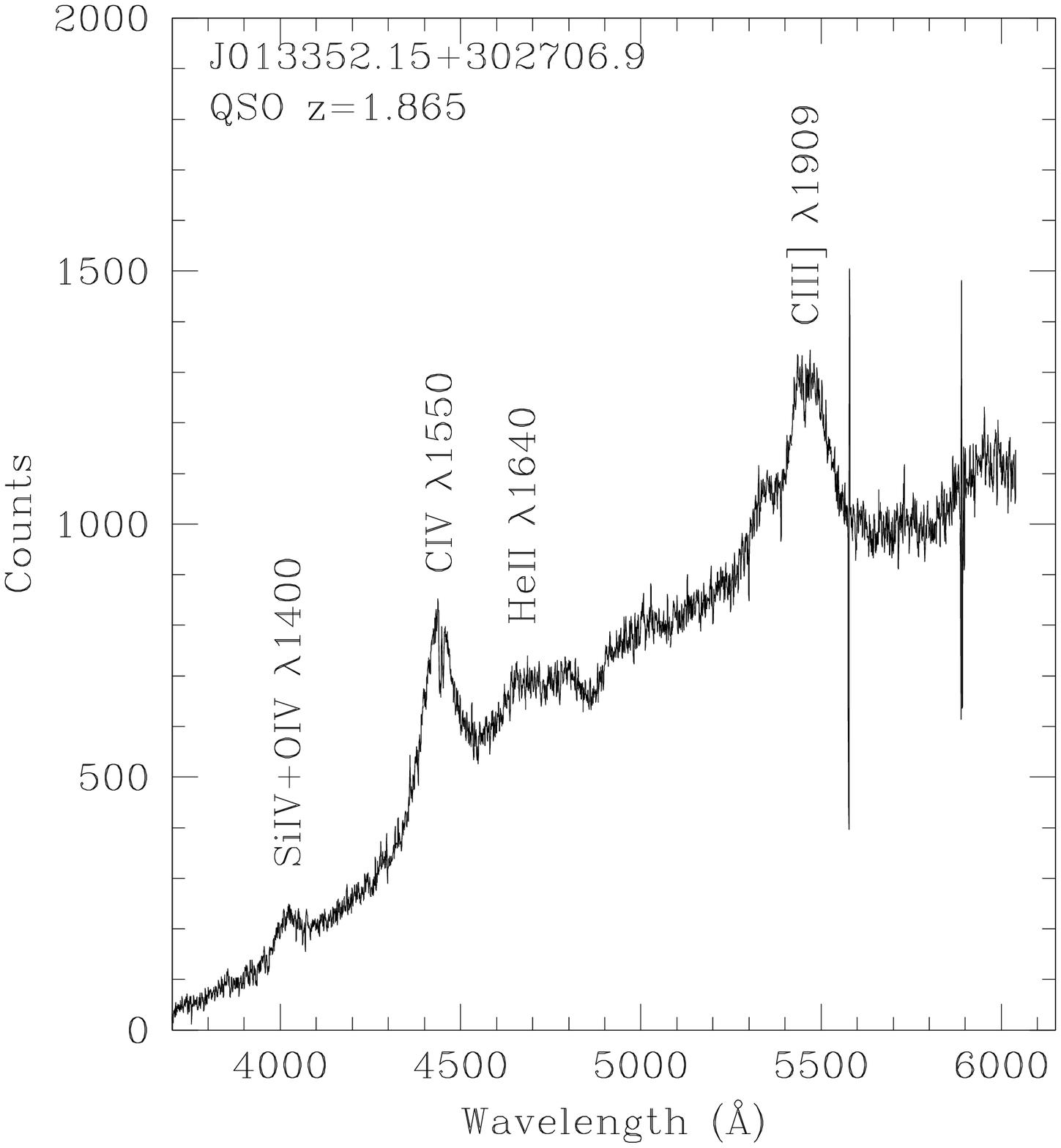}
\caption{Continued
}
\end{figure}

\clearpage

\clearpage

\clearpage

\clearpage

\clearpage

\begin{deluxetable}{l l l l }
\tabletypesize{\scriptsize}
\tablecaption{\label{tab:journal} Hectospec Observations}
\tablewidth{0pt}
\tablehead{
\colhead{LGGS}
&\colhead{UT Date}
&\colhead{Exp.\ (s)\tablenotemark{a}}
&\colhead{Grating} 
}
\startdata
J004126.23+405326.8 & 2018 Sep 14 & 9000 & 270 \\
J004130.25+412320.3 & 2018 Sep 14 & 9000 & 270 \\
J004140.65+405818.6 & 2018 Sep 14 & 9000 & 270 \\
J004141.44+411916.9 & 2018 Sep 14 & 9000 & 270 \\
J004141.50+403118.1 & 2009 Oct 15 & 7964 & 600 \\
J004145.19+412623.6 & 2018 Sep 14 & 9000 & 270 \\
J004214.98+403909.8 & 2009 Oct 15 & 7964 & 600 \\
J004543.44+412031.2 & 2011 Sep 26 & 3600 & 270 \\
J013201.68+303938.7 & 2018 Oct 6 & 9000 & 270\\
J013216.50+301607.3 & 2010 Oct 9 & 8100 & 600 \\
J013253.60+303815.2 & 2018 Oct 6 & 9000 & 270 \\
J013316.28+301752.3 &  2010 Oct 9 & 8100 & 600 \\
J013238.98+304428.9 & 2018 Oct 6 & 9000 & 270 \\
J013313.47+305709.6 & 2018 Oct 6 & 9000 & 270 \\
J013352.15+302706.9 & 2009 Oct 17 & 7740 & 600 \\
J013425.46+305159.9 & 2018 Oct 6 & 9000 & 270\\
\enddata
\tablenotetext{a}{Taken as 3 consecutive exposures}
\end{deluxetable}

\clearpage

\begin{deluxetable}{l l l l l l l }
\tabletypesize{\scriptsize}
\tablecaption{\label{tab:QSOs} Newly Found Extragalactic Objects}
\tablewidth{0pt}
\tablehead{
\colhead{LGGS}
&\colhead{Gal.}
&\colhead{$\rho$\tablenotemark{a}}
&\colhead{V\tablenotemark{b}}
&\colhead{z}
&\multicolumn{1}{l}{Image}
&\multicolumn{1}{l}{Comments}
}
\startdata
\multicolumn{7}{c}{QSOs} \\ \hline
J004126.23+405326.8 & M31&0.29 &19.75 & 1.157 &Point  & Chandra X-ray,  GALEX UV, and Spitzer IR source \\
J004130.25+412320.3 & M31&0.73 &19.55 & 1.895 & Point & BALQSO\tablenotemark{c}; XMM-Newton X-ray and WISE IR source\\
J004141.44+411916.9 & M31&0.54 &18.62 & 1.245 & Point  & XMM-Newton X-ray+UV, and WISE IR source; WISE phot.\ QSO cand.\ z=1.1\\
J004141.50+403118.1 & M31&0.87 &19.13 &  2.151 &Point & XMM-Newton X-ray and WISE IR source\\
J004145.19+412623.6 &  M31&0.70 &19.93 & 1.293 &Point & XMM-Newton X-ray+UV, Spitzer IR, and GLB\tablenotemark{d}radio source \\
J004214.98+403909.8 &  M31&0.85 &19.35 & 1.774 &Point &XMM-Newton UV and WISE IR source\\
J004543.44+412031.2 & M31&1.19 & 19.17 & 1.323 &Point &XMM-Newton X-ray+UV, and Spitzer IR source  \\
J013216.50+301607.3 & M33&1.11 & 18.94  & 1.541 &Point  &XMM-Newton X-ray, GALEX UV, and WISE IR source; SDSS phot.\ QSO cand.\ z=1.5   \\
J013253.60+303815.2 & M33&0.66 & 19.91 & 0.365 &Point & XMM-Newton X-ray+UV, and WISE IR source\\
J013316.28+301752.3 &  M33&0.75 &19.49 & 1.918 &Point  & XMM-Newton X-ray+UV, and WISE IR source; WISE phot.\ QSO cand.\ z=2.4 \\
J013352.15+302706.9 & M33&0.48 &19.44 & 1.865   &Point  & XMM-Newton X-ray+UV, and WISE IR source\\ \hline
\multicolumn{7}{c}{Galaxies} \\ \hline
J004140.65+405818.6 &M31&0.23&19.58 & 0.268 &Sl.\ extended &AF-type spectrum. Previous globular cluster cand.\tablenotemark{e}\\
J013201.68+303938.7 & M33&1.29 &18.97 & 0.188  &Extended &AF-type spectrum+neb.\ emiss. Previous RSG/foregound cand.\tablenotemark{f}\\
J013238.98+304428.9 & M33&0.94 &19.77 & 0.172  &Fuzzy and sl.\ extended &Elliptical-type spectrum. Previous globular cluster cand.\tablenotemark{g}\\
J013313.47+305709.6 & M33&0.93& 19.20 & 0.123 &Sl.\ extended &Elliptical-type spectrum+neb.\ emiss. Previous globular cluster\tablenotemark{h} \\
J013425.46+305159.9 & M33& 0.48 &19.12 & 0.135 &Sl.\ extended & AF-type spectrum+neb.\ emiss. Previous globular cluster\tablenotemark{i} \\
\enddata
\tablenotetext{a}{Projected galactocentric distance in units of the Holmberg radius, as given in the LGGS.} 
\tablenotetext{b}{Photometry from LGGS}
\tablenotetext{c}{Broad absorption line QSO, based upon blue-shifted broad absorption lines.}
\tablenotetext{d}{350 MHz source in \citealt{2004ApJS..155...89G}}
\tablenotetext{e}{[KLG2007] GC2 85 in \citealt{2007AJ....134..706K}}
\tablenotetext{f}{[Ma95a] SA 45b-44 in \citealt{MasseyRSGs}}
\tablenotetext{g}{[SSA2010] 683 in \citealt{2010ApJ...720.1674S}}
\tablenotetext{h}{[SSA2010] 1020 in \citealt{2010ApJ...720.1674S}; also X-ray source ChASeM33 68}
\tablenotetext{i}{[SSA2010] 2024 in \citealt{2010ApJ...720.1674S}}
\end{deluxetable}

\clearpage
\begin{deluxetable}{l l l l l l c }
\tabletypesize{\scriptsize}
\tablecaption{\label{tab:Previous} Previously Known QSOs with the LGGS Fields}
\tablewidth{0pt}
\tablehead{
\colhead{LGGS}
&\colhead{Gal.}
&\colhead{$\rho$\tablenotemark{a}}
&\colhead{V\tablenotemark{b}}
&\colhead{z}
&\colhead{Image} 
&\colhead{Reference}
}
\startdata
J003856.68+403446.8 &M31&0.83& 20.77 & 2.76  & Point & 1 \\
J004137.93+410107.9 & M31&0.22&19.75 & 1.45  & Point & 1\\
J004149.86+402818.1 & M31&1.00&19.72 & 1.23  &Point & 2 \\
J004457.94+412343.9& M31&0.78& 19.91 & 2.11 &Point & 3\\
J004527.30+413254.3& M31& 0.78 &19.90  & 0.22 &Point &  4 \\
J004528.26+412944.2 & M31&0.86 &19.33  & 0.20 &Point  & 5 \\
J004655.52+422050.2 & M31&0.84 &17.93  & 0.31 &Point  & 6 \\
J004742.84+421017.0 &M31&1.03 & 20.27 & 1.45 & Point  & 1 \\
J013151.09+301857.3 & M33&1.31 & 20.42 & 1.31 &Point  & 7\\
J013211.51+301417.2 &M33& 1.18 &20.08 & 1.18 & Point  & 2 \\ 
J013322.09+301651.4 & M33&0.78 & 19.42 & 2.84 &Point   &  8 \\
J013331.94+310536.8 & M33&1.09 & 18.64 & 0.79 &Point &  2 \\
J013405.80+301732.9 & M33&0.91 & 20.91 & 2.70 &Point & 7\\
J013445.02+304928.0 & M33&0.59 & 20.11 & 1.99 &Point  &  8 \\
J013525.42+310735.2 & M33&1.18 & 21.89 & 1.18 &Point  &  7
\enddata
\tablerefs{
(1) \citealt{NeugentM31};
(2) \citealt{LAMOST};
(3) \citealt{2010AA...512A...1M};
(4) \citealt{Fred}; 
(5) \citealt{2018ATel12250....1N};
(6) \citealt{2013MNRAS.432..866R};
(7) \citealt{2017AA...597A..79P};
(8) \citealt{NeugentM33}.
}
\tablenotetext{a}{Projected galactocentric distance in units of the Holmberg radius, as given in the LGGS.}
\tablenotetext{b}{Photometry from LGGS.}
\end{deluxetable}

\clearpage

\bibliographystyle{aasjournal}
\bibliography{masterbib}

\end{document}